\documentclass[showpacs,amsmath,amssymb,twocolumn,aps]{revtex4}

\usepackage{graphicx}
\usepackage{epsfig}
\usepackage{pstricks}
\usepackage{pst-plot}
\usepackage{pst-3dplot}
\usepackage{pst-plot}
\usepackage{pst-xkey}
\usepackage{xkeyval}
\usepackage{psfrag}
\usepackage{dcolumn}
\usepackage{bm}

 \def\comment#1{}
 \def\mn#1{}

\begin{document}

\title{Modelling two-dimensional Crystals with Defects under Stress:
 Superelongation of Carbon Nanotubes at high Temperatures}

\author{J\"urgen Dietel}
\affiliation{Institut f\"ur Theoretische Physik,
Freie Universit\"at Berlin, Arnimallee 14, D-14195 Berlin, Germany}
\author{Hagen Kleinert}
\affiliation{Institut f\"ur Theoretische Physik,
Freie Universit\"at Berlin, Arnimallee 14, D-14195 Berlin, Germany}
\affiliation{ICRANeT, Piazzale della Repubblica 1, 10 -65122, Pescara, Italy}
\date{Received \today}

\begin{abstract}
We calculate analytically the phase diagram of a two-dimensional
square crystal and its wrapped version 
 with defects
under external homogeneous
stress as a function of temperature using a simple
elastic lattice model that  allows for
defect formation. 
The temperature dependence turns out to be very weak.
The results are relevant
for recent stress experiments on carbon nanotubes.
Under increasing stress, we find
a crossover regime which we identify
with a cracking transition that is almost independent of temperature.
Furthermore, we find an almost stress-independent melting point.
In addition, we derive an enhanced
ductility  with relative strains before cracking between
200-400\%,
in agreement with
carbon nanotube experiments.
The specific values depend on the Poisson ratio and the
angle between the external force and the crystal axes. 
We give arguments that the results for carbon nanotubes are 
not much different to the wrapped square crystal. 
\end{abstract}

\pacs{62.20.F-, 61.46.Fg, 61.72.Lk, 64.70.dm}

\maketitle

\section{Introduction}
The discovery of macroscopical two-dimensional (2D) graphene sheets by
mechanical cleavaging \cite{Nosolev1} has demonstrated that
free-standing or suspended 2D crystals can exist despite
their large 2D positional fluctuations.
Since then,
a variety of other free-standing 2D crystallites have been
 prepared
\cite{Nosolev2}. These crystals are stabilized by
fluctuations in the solid plane
as verified
experimentally in Ref.~\onlinecite{Meyer1}, following
the predictions
in Ref.~\onlinecite{Nelson1}.
The wrapped version of the 2D free-standing graphene
had been found much earlier in 1991 \cite{Iijima1}.
Due to their high strength,
the mechanical properties of such  materials
have recently attracted great interest.

The behavior of three-dimensional (3D) crystals
as a function of
stress is well known.
For small stresses, they
expand elastically
with a linear
stress-strain curve.
Above the
yield point the curve flattens
due to the irreversible plastic deformation.
At even higher stress,
cleavage sets in
with further
fracture. If the plastic region is small
or absent, the material is called {\em brittle\/}, otherwise {\em ductile\/}.

A similar stress-strain curve
was
expected for 2D crystals or their wrapped versions.
Yakobson {\it et al.} \cite{Yakobson1}
was one of the first to determine the cracking strain
of single-wall carbon nanotubes (SWNT)
by computer simulation.
Since then, there have been
many similar studies using different
simulation methods (see \cite{Haskins1} and references therein).
Most of these
observed a cracking strain between
$15\% $ and $40\%$ depending on the chirality of the tube at room temperature
$ T \approx 300$K. The results of the different simulations
differ widely.
Experimental values  for
ropes of SWNTs found cracking strains of $ 6\% $ \cite{Walters1, Yu1} and
$ 13 \% $ \cite{Yu2} for multiwall nanotubes. Huang {\it et al.} \cite{Huang1}
measured less than $ 15\% $ for tensile failure at $ 300 $K
which is defined either by the yielding strain  for ductile
nanotubes or by the cracking strain for brittle ones.
At high temperatures they were able to
go to extreme elongations of $ 270\% $ before
cracking. Due to the large temperature, the SWNTs show an extremely
ductile behavior with kink motion along the tubes. These were
interpreted as defects which do not only perform glide but also
climb motion in the SWNT \cite{Huang2, Ding1} at high-temperatures.
In Ref.~\onlinecite{Tang1}, a molecular dynamic simulation for various
SWNTs at high temperatures was carried out exhibiting
large defect formations before cracking.

Due to large activation barriers, the strain value
of the yield point is mostly dominated by
the creation of closely lying defect pairs of opposite "charge"
forming dipoles at low
temperatures. The poles separate by
glide motions at increasing temperatures.  By calculating the energies of
Stone-Wales defects (SW) one finds, by simulation
a strain value between $ 6\%$
(arm-chair tube) and $ 12\% $ (zig-zag) \cite{Zhang2, Nardelli1,
Yakobson2}, where these defects possess negative formation energy.
The results were consistent with experiments
 based on
measuring electronic scattering in the tube \cite{Bosovic1}.
Plastic
behavior for various
 SWNTs is seen to set in at relative strains between  $ 5\%$ and $ 10\% $.

From numerical simulations,
we know that the
activation barriers for SW defects
are quite large \cite{Dumitrica1, Dumitrica2, Zhao1}.
It depends on
the time duration of
stress or heat,
how the SW defects form.
For example, brittle SWNTs show a defect formation that leads
immediately to
cleavaging, and subsequent cracking.

The purpose of this note is to study
these processes with the help of an extension of
a model
introduced in Refs.~\cite{GFCM2, Dietel1}
to describe
crystal melting of 2D and 3D lattices.
The model contains linear elastic forces
coupled minimally
to an additional integer-valued
plastic field
to allow for defect formation.
This model is here extended by
an external stress. By modifying this
we can investigate
phase transitions and instabilities of a 2D
crystal under stress at  finite temperatures.
 We shall restrict ourselves mainly 
to a square-lattice  model and its wrapped version, for simplicity.
The more realistic triangular and honeycomb lattices will
be treated in the future by extending the corresponding melting models
\cite{Dietel1}. 
The basic physics
will not be much different for different lattice symmetries.
We come back to this point in Sect.~VI below where we 
give more arguments that also quantitatively it should 
not change much going from square crystals to 
triangular and honeycomb lattices and its wrapped versions.  
   
The main advantage of square lattices is that one can
easily calculate
partition functions,
which require sums over
integer-valued defect fields. The sums
are simplest for square crystals with
small Poisson ratios
\cite{GFCM2}. It turns out that SWNTs and graphene
are systems with small Poisson ratio $ \sim 0.14$, making them
well suited for applying this technique.
The sums can be performed with a technique
developed for XY models of superfluidity,
using an so-called inverse Villain (iV) approximation \cite{GFCM2}.
The defect model in this approximation will briefly be referred to as
{\em cosine model\/}.
In the cosine model,
the defect aspects can treated by mean-field methods.

In the following, we will first discuss the phase diagram of extended 
2D square crystals starting with the phase diagram of
the cosine-model in mean-field approximation. We shall
 find
a second-order phase transition line which is
 identified as the cracking
transition connecting
 the melting point with a point at zero temperature.
In addition, we encounter a vertical second-order transition line
at constant temperature starting at the melting point.\mn{check}

Next, we discuss the full theory without
the iV-approximation. We shall find a similar phase diagram
where now the second-order cracking line in the iV-approximation
is almost everywhere a crossover.
For temperatures near the melting transition our theory give
relative strains  of 200-400\% before cracking.
This is in accordance to the high-strain values of the experiments of
Huang {\it et al.} \cite{Huang1, Huang2} for SWNTs.
We find extended defect configurations  
before cracking consisting 
of homogeneously distributed defect stripes. 

Finally, we will discuss the physics of large wrapped square crystals 
under stress. 
We find the same phase diagram, stress-strain function and  
cracking stress for the wrapped version of a square crystal as for 
the 2D extended crystal. The main difference lies in the fact that for 
{\it achiral} tubes  which we define by the property that the vector 
along the circumference of the square tube lies not in the direction 
of a crystal axis show spiral-like defect configurations under stress. 
In accordance to the experiments \cite{Huang2} defect glide and climbs are 
relevant in tubes.

We point out
that
within our theory
 it is impossible
to find the correct yield
point at small temperatures where plasticity sets in, since
our model does not
 really account for
the
true activation barriers
 \cite{Dumitrica1, Dumitrica2, Zhao1}.
Since activation energies at high-temperatures are no longer
relevant because defects overcome the barriers by
thermal fluctuations, we expect that the yielding point
tends to zero stress leading to an extensive dislocation creep
seen in the experiments of Huang {\it et al.} \cite{Huang1}.
This is the temperature regime where our theory gives the
correct phase diagram.

The paper is organized as follows. In Section II
we state the model. Section III contains the calculation of the phase
diagram within the iV-approximation in mean-field. In Section IV,
we discuss the full crystal Villain model. Section V contains a discussion
of the true phase diagram for a square crystal under stress
by taking into account the discussions in  Section III and IV.
We also discuss in this section the cracking stress and the
relative strains
 as a function of the external stress before cracking.
In Sect. VI, we discuss the modifications of our results when 
considering wrapped versions of 2D crystals and carbon nanotubes

\section{Model}
The partition function used here for the square crystal
was proposed in Ref.~\onlinecite{GFCM2}.
It can be written in the canonical form as
a functional integral
\begin{equation}
Z_{\rm fl}\!= \! \int  {\cal D}[u_i,\sigma_{ij},n_{ij}] e^{-
 \left(H_{d}[u_i,\sigma_{ij},n_i]+H^1_{\sigma^0}[u_i]\right)/k_BT
} \,,  \label{1}
\end{equation}
where
\begin{eqnarray}
\frac{ H_{d}[u_i,\sigma_{ij},n_{ij}]}{k_BT}  &=&
 \sum_{{\bf x}}\Bigg\{ \frac{1}{2 \beta}
 \Bigg[\frac{1}{2} \sigma _{ij}^2
 -
 \frac{\lambda}{4(\lambda+\mu)}
\Big(
\frac{\overline{\nabla}_i}{\nabla_i}  \sigma _{ii}\! \Big)^2
 \Bigg]  \nonumber       \\
&&~~~~~~~~~~~-
2 \pi i
  \sigma_{ij}\left(
  \nabla_{i}  u_j +
  n_{ij}\right)\Bigg\} ,         \label{2}
 \end{eqnarray}
and
\begin{equation}
 H^1_{\sigma^0}[u_i]
= - v_F \sum_{{\bf x}}
\sigma^0_{ij} \nabla_i u_j \,.     \label{3}
\end{equation}
Here $ v_F=a^2 $ is the area  of the fundamental cell where $ a $ is the
lattice constant.
The exponent in Eq.~(\ref{1})
contains the canonical representation
of elastic and plastic energies,
summed over the lattice sites
$ {\bf x} $ of a 2D lattice. The canonically conjugate
variables of the distortion fields
$\nabla _ju_{i}$ are the stress fields
 $ \sigma_{ij} $ for $ i \le j $ with the abbreviation
$ \sigma_{21} \equiv \sigma_{12}$
 \cite{GFCM2}. The stress field $ \sigma^0_{ij}  $ accounts for
  external forces
applied to the boundary of the crystal.
The parameter $ \beta $
 is
proportional to the inverse temperature,
 $ \beta \equiv
a^2 \mu /k_B T (2\pi)^2  $.

The integer-valued fields $ n_{ij}({\bf x}) $ in Eq.~(\ref{2})
are defect gauge fields  representing the
jumps of the displacements field $u_i({\bf x})$ over the {\it Volterra surfaces}.
The lattice derivatives $ \nabla_i $ and their conjugate counterparts
$ \overline{\nabla}_i $
denote
 lattice differences for a cubic 2D crystal.
In Eq.~(\ref{2}),  the defect gauge fields $ n_{ij} $ is coupled
 minimally to the displacements fields $ u_i $. Note that we
do not have this minimally coupling in the stress term
$H^1_{\sigma^0}[u_i]
 $ because the external force only acts on the surface
of the crystal.

The measure of functional integration in (\ref{1}) is
\begin{eqnarray}\!\!\!\!\!\!\!\!
\!\!\!\!\!\!\!\!
\!\!\!\!\!\!\!\!
\!\!\!\!\!\!\!\!
\!\!\!\!\!\!\!\!
 \int {\cal D}[u_i,\sigma_{ij},n_{ij}]\!=\!
\left[\frac{\mu}{4(\lambda+\mu)}\right]^{N/2}
\!\!\left[\frac{1}{2\pi\beta}\right]^{3N/2}~~~~~~~~~~~\nonumber \\
~\times \!
\left\{\! \prod_{{\bf x}}\!
 \Bigg[ \prod_{i\leq j} \int_{-\infty}^\infty d\sigma_{ij}\!\Bigg]
 \Bigg[\!\prod_{ij}\sum_{n_{ij}({\bf x})=-\infty}^{\infty} \Bigg]
 \Bigg[\!\int_{-\infty}^\infty\!\frac{d {\bf u}}{a}
 \Bigg]\! \right\}, ~ \label{20}
\end{eqnarray}
where $ N $ is the number of lattice sites.

Let us integrate out
the stress fields $ \sigma_{ij} $
in (\ref{1}).
 This leads to the partition
function of an elastic Hamiltonian with a minimally defect gauge
field under stress \cite{GFCM2}.
We use {\it free boundary conditions}  for the crystal.
These are taken into account
by  separating the displacements field integration in the partition function
over zero momentum terms  $ u_i(q=0) $ and terms
with $ u_i(q\not=0) $ \cite{Nelson2}. In the following we first integrate
out the zero momentum displacement fields $ u_i(q=0) $. One should
now take care of this integration due to the following fact:
A crystal which is homogeneously deformed has three
independent strain directions instead of two which is suggested by
the counting of the number of displacement fields.
One can take care of this by integrating over the three independent
strain fields
$ u_{ij}=(\nabla_i u_j +\nabla_j u_i)/2 $ for $ q=0 $ \cite{Nelson2}
instead of the displacement fields $ u_i(q=0) $.\mn{dont understand your counting} This leads
to the partition function
\begin{align}
&\!\!\!\!\!\!\! Z_{\rm fl}\!= \! \int  {\cal D}[u_i,\sigma_{ij},n_{ij}]
 \label{22} \\
& \times \exp\left[-
 \frac{\left(H_{d}[u_i,\sigma_{ij},n_i]+\tilde{H}^1_{\sigma^0}[u_i]+
\tilde{H}^2_{\sigma^0} \right)}{k_B T}
\right] \,,  \nonumber
\end{align}
with
\begin{align}
 \tilde{H}^1_{\sigma^0}[u_i]
&=  v_F \sum_{{\bf x}}  \sigma^0_{ij}\left(
  \nabla_{i}  u_j +
  n_{ij}\right)   \,,     \label{23} \\
  \tilde{H}^2_{\sigma^0}
&= -\frac{v_F}{\mu} \sum_{{\bf x}}
 \Bigg[\frac{1}{4} \left(\sigma^0_{ij}\right)^2
 -  \frac{\lambda}{8(\lambda+\mu)}
\Big(\sigma^0_{ii}\! \Big)^2
 \Bigg]              .    \label{24}
\end{align}
The  displacements fields $ u_i({\bf x}) $ of
non-zero momentum are integrated out
in the integration measure (\ref{20}) with
{\it periodic
boundary conditions}. Because of this,
the first term in
$ \tilde{H}^1_{\sigma^0}[u_i]$
is  actually zero for homogeneous external
stress. It is only displayed in
(\ref{23})
 to exhibit
the minimal coupling
nature of the defect fields in (\ref{22}).
The Hamiltonian
$ \tilde{H}^2_{\sigma^0} $ 
describes
the well-known elastic
energy of a 2D crystal under a constant stress
if no defects are present.
Due to its similarity with the
Villain-model of superfluidity \cite{GFCM2},
 the model
(\ref{22}) will be called the {\em Villain model of crystals\/}.

\section{Square Crystal in Cosine Model}
In the following, we restrict our attention
to an external homogeneous stress
along the x-axis, i.e.
$ \sigma^0_{ij} = \sigma^0 \delta_{i 1}
\delta_{j 1} $.
We shall  calculate the partition function (\ref{1}) in the
iV-approximation in mean-field for $ \nu=0 $.
This was done in the case of zero external stress in the
textbook \cite{GFCM2}.

We now describe the procedure when taking into account
external stresses.
First, we integrate out in (\ref{1})
the stress fields $ \sigma_{ij} $.
Then one can sum in the iV-approximation over the integer
defect fields by restricting
the displacements fields to the fundamental cell \cite{GFCM2}.
This leads up to structural factors being a function of
a parameter
 $ \beta_{\bar V} $
related to the parameter $ \beta $ by an inverse Villain transformation
\cite{GFCM2}
\begin{equation}
\beta = -\frac{1}{2 \ln[I_d(\beta_{\bar V})]},     \label{42}
\end{equation}
where
$ I_d(\beta) $ is defined by
$ I_d(\beta)= I_1(\beta)/ I_0(\beta) $ and $  I_0,  I_1 $ are modified
Bessel functions of the first kind.
Over the relevant regime
treated in this paper, $ \beta_{\bar V}$
is roughly
 proportional to $ \beta $ \cite{GFCM1,GFCM2}.

The properties of the model (\ref{1}) can then be calculated
 approximately from the
lattice partition function
\begin{align}
& Z_{\rm fl}\sim
\prod_{{\bf x}}\!
 \Bigg[\int\limits_{-a/2}^{a/2}\frac{d {\bf u}}{a}
 \Bigg]  \exp \Bigg[\!
- \beta_{\bar V}  H^{XY}_{d}-2 \tilde{H}^2_{\sigma^0}/k_B T
  \Bigg] \,, \label{30}
 \end{align}
with the cosine Hamiltonian
 \begin{align}
&  H^{XY}_{d}  = -  \sum_{x}
\cos\left[2 \pi( \nabla_1 u_2 + \nabla_2 u_1)\right] \nonumber \\
& + 2 \cos\left[ 2 \pi(
\nabla_1 u_1+ \frac{1}{2 \mu} \sigma^0)\right] +
2 \cos\left[ 2 \pi
\nabla_2 u_2\right]  \,. \label{40}
\end{align}
The displacements fields
satisfy periodic boundary conditions.
The Hamiltonian (\ref{40}) represents two
one-dimensional (1D) XY-models which are coupled by the first term.
This coupling term causes the melting transition for
$ \sigma^0 =0 $  \cite{GFCM2}.

There are two identical ways to derive a mean-field approximation
from the partition function (\ref{40}) \cite{GFCM2}.
Either one  uses the Bogoliubov variation principle
with a trial Hamiltonian,
 or one inserts
constraint fields leading to the variational mean-field free energy
$ f_{\rm var} $ per atom in the lattice
\cite{GFCM2}. In the following we discuss
the first way. As a trial partition function we use
\begin{equation}
Z_0= \prod_{\bf x,i} \int\limits_{-a/2}^{a/2} \frac{du_i({\bf x})}{a}  \exp\left[\alpha_i
\cos\left(2 \pi \frac{u_i({\bf x})}{a}\right)\right].        \label{43}
\end{equation}
In general $ \alpha_1 \not= \alpha_2 $ for an external stress which is
not zero. In Ref.~\onlinecite{GFCM2} one uses the same Ansatz for the  trial
partition function where $ \alpha_1 = \alpha_2 $ when the external
stress is zero.
By using Peierls inequality we obtain an upper bound
for the actual free energy per lattice site $ f_{\rm fl}=
- k_B T\ln(Z_{\rm fl}) /N  $
\cite{GFCM2} given by $  f_{\rm var} $ with
\begin{align}
& f_{\rm var}= - k_B T \bigg\{\log[I_0(\alpha_1)]+
\log[I_0(\alpha_2)]  \label{50} \\
&
+ \beta_{\bar V} I_d(\alpha_1)^2
I_d(\alpha_2)^2 + \beta_{\bar V}
2  \cos\left(\frac{\pi \sigma^0}
{\mu}\right)   I_d(\alpha_1)^2   \nonumber  \\
& + \beta_{\bar V}
2  I_d(\alpha_2)^2  - \alpha_1 I_d(\alpha_1)
 -  \alpha_2 I_d(\alpha_2) \bigg\}- \frac{v_F(\sigma^0)^2}{2 \mu} \,.
\nonumber
\end{align}
The best approximation for $ f_{\rm fl} $
is given by the minimum
 of $ f_{\rm var} $ with respect to $ \alpha_1 $, $ \alpha_2 $.
We mention that the first term in (\ref{40})
corresponds to the third term in (\ref{50}).

\begin{figure}
\psfrag{x}{$1/\beta $}
\psfrag{y}{$\frac{\sigma^0 }{2 \mu}$}
\psfrag{A}{$ \frac{\sigma^0_b}{2\mu} $ }
\psfrag{B}{$1/\beta_m$}
\psfrag{R1}{${\cal R}_1$}
\psfrag{R2}{$ {\cal R}_2$}
 \psfrag{R3}{$ {\cal R}_3$}
 \includegraphics[height=6cm,width=7.5cm]{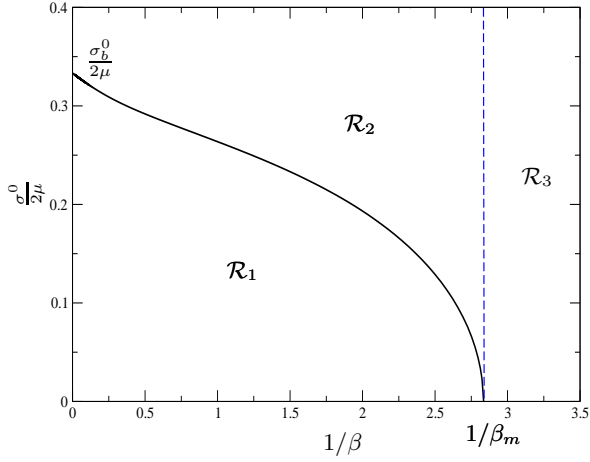}
\caption{We show the phase diagram in the $ (1/\beta,\sigma^0/2 \mu) $-plane of
a 2D crystal under stress $ \sigma^0 $ calculated in the
mean-field iV-approximation by using (\ref{60}), (\ref{70}) and (\ref{80}).
The intersection of the transition lines
and the temperature axis denoted by $ 1/ \beta_m $ is the
melting point for zero external stress.}
\end{figure}

The analysis of the saddle point equations for (\ref{50}) is straight forward.
We summarize in the following the results briefly.
We find four distinct solutions characterized by $ \alpha_i=0 $ or
$ \alpha_i \not=0 $ for $ i=1,2 $.
Only three of these saddle point
solutions are minima of $ f_{\rm var} $ in certain $(T,\sigma^0)$ regimes
corresponding to phase regions.
To be more specific, one  can show that the saddle point with
 $ \alpha_1 \not= 0$, $ \alpha_2 = 0$ is not a minimum of $ f_{\rm var} $
when comparing its free energy value with the other saddle point values.
One obtains the following three different $ \beta_{V} $
 regimes $ {\cal R}_i$ as a function of $ \sigma^0 $:
\begin{eqnarray}
 {\cal R}_1 & : & \beta_{\bar V} >\beta_{\bar{V},b}(\sigma^0)  \,,        \nonumber \\
 {\cal R}_2 & : &  1/2 \le \beta_{\bar V} \le \beta_{\bar{V},b}(\sigma^0) \,,
 \nonumber  \\
 {\cal R}_3 & : &   1/2 > \beta_{\bar V}  \,.  \label{60}
 \end{eqnarray}
The function $ \beta_{\bar{V},b}(\sigma^0) $ is defined by the implicit equations
\begin{eqnarray}
\beta_{\bar{V},b} & = & \frac{\alpha}{4 I_d(\alpha)} \,,  \nonumber \\
\cos\left(\frac{\pi \sigma^0}{\mu}\right) & = & \frac{2}{\alpha}
I_d(\alpha)- \frac{1}{2}I^2_d(\alpha) \,. \label{65}
\end{eqnarray}
The values of $ \alpha_i $ in the regimes $ {\cal R}_i $ are
\begin{eqnarray}
{\cal R}_1 & : &
I_d(\alpha_1) =
\sqrt{\frac{1}{2 \beta_{\bar V}} \frac{\alpha_2 }{I_d(\alpha_2)}-2} \,,  \nonumber \\
& & I_d(\alpha_2) =
\sqrt{\frac{1}{2 \beta_{\bar V}} \frac{\alpha_1}{I_d(\alpha_1)}-2
\cos\left(\frac{\pi \sigma^0}{\mu }
\right)}    \,,  \nonumber \\
{\cal R}_2\!\!& : &\!  \alpha_1=0 \quad , \quad
4 \beta_{\bar V}  = \frac{\alpha_2}{I_d(\alpha_2)} \,, \nonumber \\
{\cal R}_3\! & : & \! \alpha_1=0 \quad , \quad  \alpha_2=0 \,. \label{70}
\end{eqnarray}
The free energies in the various regimes are given by
\begin{eqnarray}
 {\cal R}_1 \!&  : & \!   f_{\rm var}=
 -k_B T \bigg\{\! \!  \log[I_0(\alpha_1)]\! + \!\log[I_0(\alpha_2)]\!- \!\frac{\alpha_1}{2} I_d(\alpha_1)
  \nonumber \\
& &  \!   - \frac{\alpha_2}{2} I_d(\alpha_2)
\! -\!  \frac{\alpha_2}{2}
\frac{I_d(\alpha_1)^2 I_d(\alpha_2)}{\left(2 + I_d(\alpha_1)^2\right)}
 \bigg\} \!- \!\frac{v_F (\sigma^0)^2}{2 \mu},
     \nonumber \\
 {\cal R}_2  \!& : & \!
 f_{\rm var} = -k_B T \bigg\{\! \log[I_0(\alpha_2)]
-  \frac{\alpha_2}{2 } I_d(\alpha_2)\! \bigg\}\! -\! \frac{v_F (\sigma^0)^2}{2 \mu}  \,,
\nonumber \\
 {\cal R}_3 \! & : & \!
  f_{\rm var}= - \frac{v_F(\sigma^0)^2}{2 \mu} . \label{80}
\end{eqnarray}

In Fig.~1, we show the phase diagram calculated with the help
of (\ref{42}), (\ref{60}), (\ref{70}) and (\ref{80}).
We obtain two second-order phase
transitions between the three regions $ {\cal R}_i $.
The fact that the transition
between ${\cal R}_1 $ and ${\cal R}_2 $ is of second-order type can be best
seen by using the stationarity condition for $ f_{\rm var} $ in
the saddle point.
This transition corresponds to the cracking transition.
The transition line intersect the $ 1/ \beta $-axis at the
melting transition point $ 1/ \beta_m $ corresponding to
$ 1/\beta \approx  2.85 $.
On the low-temperature side,
the transition line intersects the $ \sigma^0/2\mu $-axis at
$ \sigma_b^0/2\mu =1/3$.
We note
that in the regime $ {\cal R}_2 $ where $ \alpha_1 =0 $
we obtain from  (\ref{50}) the mean-field variational energy
of a 1D XY-model. The second-order transition between $ {\cal R}_2 $
and $ {\cal R}_3 $ corresponds then to the phase transition of a
1D XY-model obtained by the help of the mean-field approximation.
In an exact treatment of the 1D XY-model this transition is of course
not existent being only an artefact of the mean-field approximation
\cite{Joyce1}.
There is in fact an argument
that this  phase transition for the 2D crystal under stresses
is existent because for real physical systems there
exist a melting transition beyond the cracking of the crystal.
Note that we cannot get the cleavaged cracked state exactly within
our model since in real physical systems the time scale beyond cracking
are so long that the thermodynamical average is no longer fulfilled for
a concrete system. Within our formalism, we can only describe the
physics of the Gibb's state including a thermodynamical average.
In the next section we shall discuss the full model (\ref{1}) with (\ref{2})
were we still find a melting transition independent of the external stress
$ \sigma^0 $.

\begin{figure}
\psfrag{x}{$\frac{\sigma^0}{2 \mu} $}
\psfrag{y}{$\frac{\Delta u_\parallel}{a}$}
\psfrag{a1}{\scriptsize $2.5 $ }
\psfrag{a2}{\scriptsize $2.0 $ }
\psfrag{a3}{\scriptsize$1.5$}
\psfrag{a4}{\scriptsize$1.0 $}
\psfrag{a5}{\scriptsize$ 0.5 $}
\psfrag{a6}{\scriptsize$0.15 $}
\psfrag{a7}{\scriptsize$0. $}
 \includegraphics[height=6cm,width=7.5cm]{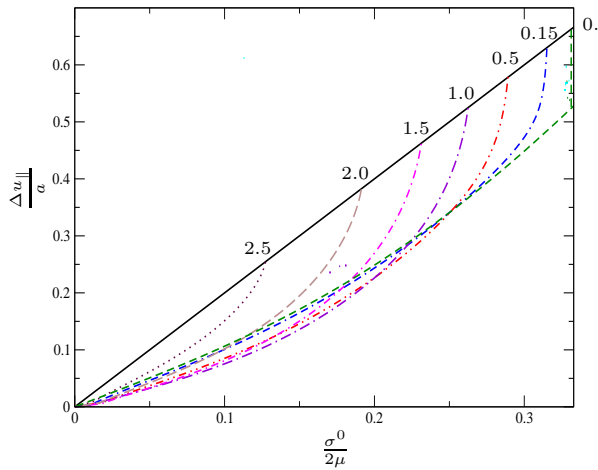}
\caption{We show the relative elongation rates $ \Delta u_\parallel /a$
in the direction of the external force as a function of the dimensionless
external stress $ \sigma^0/2 \mu $. The (black) straight
solid curve corresponds to two times the elastic  elongation, i.e.
$ 2 \sigma^0/2 \mu $. We show the elongation rates for various different
dimensionless temperatures $ 1/\beta $ shown as numbers located
at the intersection point of the
corresponding curve with the (black) solid
straight curve.
The x-axis value of the intersection point
is then given by the dimensionless cracking stress.}
\end{figure}

Next, we calculate the relative strain parallel to the external force
within the mean-field approximation of the cosine-model. It is given by
$ \Delta u_\parallel/a = \partial f_{\rm var}/ \partial (v_F \sigma^0) $.
In Fig.~2 we show relative strain values
for various dimensionless temperatures $ 1/ \beta $
as a function of the external dimensionless stress
$ \sigma^0/2 \mu $. The solid (black) curve shows
two times the elastic part
of the total relative strain, i.e. $ 2 \sigma^0/2 \mu $.
We do not obtain from the
mean-field approximation of the cosine-model elongation rates before breaking
of a few hundred percent seen in carbon nanotube experiments \cite{Huang1}
but only less than $ 2/3 $. The reason lies presumably in the mean-field
approximation which does not take into account the defect
degrees of freedom correctly. This will be shown in the next section
by taking into account
the stress degrees of freedom in the full crystal Villain model
exactly when calculating the free energy.

\section{Square crystal in Villain model}
Next, we discuss the partition function (\ref{22}) of
the full model (\ref{23}), (\ref{24}). The melting line of the 2D square
crystal  under the influence of homogeneous stress
will be calculated similarly to the stress free system
\cite{GFCM2,Dietel1}. This was done by intersecting the high and
low-temperature expansion of the partition function (\ref{22}).

First, we calculate the simpler case of the high-temperature
expansion. This was carried out formerly in Ref. \onlinecite{GFCM2}
for the case $ \sigma^0_{ij}=0 $.
We start by integrating out in (\ref{22})
the  displacements fields $ u_i $ and afterwards
the stress field $ \sigma_{ij} $. This leads us to
the high-temperature limit of the partition function
$ Z$
\begin{equation}
Z^{T \rightarrow \infty}  =
Z_{0}^{T \rightarrow \infty} e^{-2\tilde{H}^2_{\sigma^0}/k_B T}
Z_{\rm stress}                  \label{140}
\end{equation}
with
\begin{align}
&\!\!\!\!\!\!\!\!\!\!\! Z_{\rm stress}
 = \prod_{{\bf x}}  \left[
\sum_{\chi({\bf x}) \in \,{ \mathbb Z }}    \right]  \,.
   ~~~~~   \label{150}
      \\
& ~~\, \times \exp\left[-\frac{1}{4 \tilde{\beta}}
 \sum_{{\bf x},{\bf x}'} \chi({\bf x})
a^{-4} v^{-2} ({\bf x}-{\bf x}')
 \chi({\bf x}') \right]   \nonumber
\end{align}
where
$v ({\bf x}) $ is short for
$ \triangle^{-1}({\bf x}) $ and $  \triangle({\bf x}) $ is the 2D lattice
Laplacian. $ \tilde{\beta} $ is defined by
$ \tilde{\beta} \equiv  \beta (1+ \lambda/(2 \mu +\lambda)) $
and we shall use further below
the abbreviation $ \tilde{\mu} \equiv  \mu(1+ \lambda/(2 \mu +\lambda))  $.
The lowest-order approximation
to the high-temperature partition function $ Z^{T \to \infty}_0 $ is given
by \cite{Dietel1}
\begin{equation}
Z^{T \to \infty}_0
\left(2 \pi \beta \right)^{-3N/2}
\left[\frac{\mu}{4(\lambda+\mu)}\right]^{N/2}  \,.
\end{equation}
Taking only into account the dominant stress configuration
$ \chi({\bf x})= \pm \delta_{{\bf x}, {\bf x}_0} $
we obtain \cite{Dietel1}
\begin{equation}
Z_{\rm stress} \approx
\exp[2 N e^{-5/ \tilde{\beta}}] \,.
\label{160}
\end{equation}
Note that we obtain an agreement between the free energy density in the
high-temperature phase $ {\cal R}_3 $ (\ref{80})
for the cosine-model with the $ \sigma^0 $-part of the high-temperature
free energy density calculated from (\ref{140})
which is given by $ 2  \tilde{H}^2_{\sigma^0}/N $.

 Next, we calculate the low-temperature expansion. For
$ \sigma^0_{ij}={\rm const.} $ one can skip the first term in
(\ref{23}) (we left this additional vanishing term only in (\ref{23})
to show the minimal coupling form and thus the gauge degrees of
freedom of the defect fields). First, we integrate out the
displacement fields $ u_i $ and afterwards the stress-fields
$ \sigma_{ij} $. Then one obtains the partition function
of an elastic crystal under stress times a defect dependent term denoted
by $ Z_{\rm def} $
\begin{equation}
 Z^{T \rightarrow 0}  =  Z^{T \to 0}_0
e^{-\tilde{H}^2_{\sigma^0}/k_B T}                \label{170}
Z_{\rm def}
\end{equation}
with the lowest-order result \cite{GFCM2, Dietel1}
\begin{equation}
Z^{T \to 0}_0
  =  (2 \pi\beta)^{-N} \left(\frac{\mu}{\lambda+2 \mu
}\right)^{N/2}
e^{ -N \ell}           \label{200}
\end{equation}
where $ \ell \approx 1.14 $.
The defect part of the low-temperature partition function is given by
\begin{equation}
 Z_{\rm def}   = \sum_{\cal S}
  \sum_{n_{ij} \in {\cal S}}
\exp\Bigg[-\frac{1}{k_B T} (H_{\rm def}[n]+ H_{\sigma^0 }[n])
\Bigg] ,  \label{250}
\end{equation}
with
\begin{align}
& \frac{H_{\rm def}[n]}{k_B T} = 4 \pi^2 \tilde{\beta}
\sum_{{\bf x}, {\bf x}'}
\left[\epsilon_{i i'}  \nabla_i \epsilon_{j j'}
\nabla_{j} n_{i'j'}({\bf x}) \right]  \nonumber  \\
&
\qquad \qquad \times  v^2({\bf x}-{\bf x}') \left[\epsilon_{k k'} \nabla'_{k}  \epsilon_{l l'} \nabla'_{l}
 n_{k'l'}({\bf x}') \right] \,,
   \label{260}   \\
& H_{\sigma^0 }[n] =
+v_F \sum_{\bf x} \sigma^0_{ i j}({\bf x})
n_{i j}({\bf x})  \,.  \label{270}
\end{align}
The symbol $ {\cal S} $ denotes
 the set of gauge-inequivalent defect configurations $ n_{ij} $
on the lattice which omit
all
gauge-equivalent versions
 $   n'_{ij}=
 n_{ij}({\bf x})+ \nabla_i \lambda_j  $
with periodic functions $ \lambda_i $.
The first term in (\ref{250})
is a pure defect interaction energy term, the second term
a stress-defect interaction term. The latter results in the well
known Peach-Koehler force when calculating the
force on a defect configuration $ n_{ij} $ due to the external
stress \cite{Hirth1}.

In the following, we choose $ \sigma^0_{ij} = \sigma^0 \delta_{i 1}
\delta_{j 1} $ as in the last section, and
 consider the following defect configuration
\begin{equation}
n^l_{ij}({\bf x})=\pm \delta_{i,1} \delta_{j,1}
\prod^l_{m=0} \delta_{{\bf x}, a m{\bf e_2}} \,.    \label{280}
\end{equation}
From (\ref{260}) we find that this defect configuration has
an energy dependence
$ H_{\rm def}[n^l]/k_B T \sim \ln(l)  $. On the other hand the stress-defect
energy term $ H_{\sigma^0}[n^l] $ is
proportional to the length of the defect line $ l $, i.e.
$ H_{\sigma^0}[n^l]/k_B T \sim \pm l  $.
This  means that for an infinite crystal one can construct
localized defect configurations with arbitrarily small energies.
This leads us to the conclusion that for high-temperatures where
activation energies are no longer relevant, the yield point
where plasticity sets in lies almost at zero stress.

The defects of the type $ n^l $
make the defect partition sum
$ Z_{\rm def} $ in (\ref{250}) diverge, when
performing the sum
over all defect configurations in the free energy, ordered by their
geometrical size in the well-known
{\em cluster expansion\/}.
This is in contrast to the convergent sum
if one calculates
$ Z_{\rm def} $ for  $ \sigma^0= 0 $ (\ref{250})
 due to the fact
that large defect configurations have generally large
defect energies $ H_{\rm def} $ \cite{GFCM2, Dietel1}
and thus small Boltzmann factors.
Mathematically speaking, one cannot interchange the cluster expansion sum
of  (\ref{250}) with the thermodynamic limit $ N \to \infty $ for
$ \sigma^0 \not =0 $.

In order to calculate the free energy,
we define the following defect
fields
\begin{equation}  \!\!
\tilde{n}_{ij}({\bf x}) = \sum\limits_n\left( n_1 \,\delta_{x_1, a d_1 n} \,
 \delta_{i,1} \delta_{j,1}\! +
 n_2 \, \delta_{x_2, a d_2 n} \,
\delta_{i,2} \delta_{j,2} \right) .            \label{290}
\end{equation}
The integer-valued numbers $  n_i$, $ d_i $ are defined below.
We now Fourier-transform the defect fields $ \tilde{n}_{ij}({\bf x}) $
leading for $ N \to \infty $ to
\begin{align}
&  \tilde{n}_{ij}({\bf k}) =  \sum_{\bf x} \tilde{n}_{ij}({\bf x})
e^{-i {\bf k} {\bf x} }                     \label{300}      \\
&=   (2 \pi)^2  \sum_{n_1, n_2}   \delta_{i,1}\, \delta_{j,1} \,
\delta(a d_1 k_1-2 \pi n_1)
\delta(a k_2-2 \pi n_2)      \nonumber \\
&  ~~~~~~~~~~~~+ \delta_{i,2} \, \delta_{j,2} \,  \delta(a k_1-2 \pi n_1)
\delta(a d_2 k_2-2 \pi n_2)
\nonumber
\end{align}
In the following, we substitute  $ n_{ij}
\rightarrow n_{ij} + \tilde{n}_{ij} $
in Eq.~(\ref{250}). The aim is to
choose $  n_i$, $ d_i $ in such a way that the terms proportional
$ \Sigma_{\bf x}   \sigma^0_{ij} n_{ij} $ vanish in the exponent of
(\ref{250}).
From (\ref{260})  we obtain only non-vanishing contributions of the
$ n_1$, $n_2 $-sum in (\ref{300}) for $ n_1=n_2=0 $.
In order to obtain the correct result we have to be rather careful
taking the zero momentum limits in (\ref{260}). This should be done for
a finite lattice system with $ N $ vertices. In order to get the zero
momentum limit of (\ref{260}) we should first take into account that in
(\ref{22}) the integration over the zero momentum elongations $ u_i $
are excluded since this corresponds to a translation of the solid.
By carrying out the integration over the stress fields $ \sigma_{ij} $
we obtain the zero momentum limit of $ H_{\rm def}[n] $
\begin{align}
& \frac{H_{\rm def}({\bf q}\to 0)}{k_B T }= \frac{v_F}{N}
\sum\limits_{{\bf x}, {\bf x}'}
\frac{1}{2} \, \mu \, n_{12}( {\bf x})  n_{12}( {\bf x}') \nonumber \\
& ~~~~~~~~~~~~~~ +
\mu n_{ii}( {\bf x}) n_{ii}( {\bf x}')
+ \frac{\lambda}{2} n_{ii}({\bf x}) n_{jj}({\bf x}')  \label{310}
\end{align}

With the help of this expression, we can rewrite $ Z_{\rm def} $ for
$ N \gg 1  $ as
\begin{align}
& Z_{\rm def}   = \sum_{\cal S}
  \sum_{n_{ij} \in {\cal S}}
\exp\Bigg[-\frac{1}{k_B T} (H_{\rm def}[n+\tilde{n}]+
H_{\sigma^0 }[n+\tilde{n}]) \Bigg]   \nonumber \\
&   = \sum_{\cal S}
  \sum_{n_{ij} \in {\cal S}}  \exp\Bigg[-\frac{H_{\rm def}[n]}{k_B T}  \Bigg]  \cdot \exp\Bigg[-4\pi^2 \tilde{\beta} N \frac{n_1}{d_1}
\frac{\sigma^0}{\tilde{\mu}} \Bigg]  \nonumber \\
&
~~~~~~~~\times  \exp\Bigg[ 4 \pi^2 \tilde{\beta}  N \bigg\{-
\left[\left(\frac{n_1}{d_1}\right)^2 +\left(\frac{n_2}{d_2}\right)^2 \right]
  \nonumber \\
&
 \qquad
 \qquad
 \qquad
-
\frac{\lambda}{2\mu } \left(\frac{n_1}{d_1}+
\frac{n_2}{d_2}\right)^2\bigg\}
  \Bigg]
\label{320}
\end{align}
where $ {n_i}/{d_i} $ are  determined by the equations
\begin{eqnarray}
(2 \mu +\lambda) \, \frac{n_1}{d_1}  + \lambda\, \frac{n_2}{d_2}   & = &
-\sigma^0          \,, \nonumber     \\
(2 \mu +\lambda) \,\frac{n_2}{d_2}  + \lambda \, \frac{n_1}{d_1} & = & 0 \,.
\label{340}
\end{eqnarray}
These equations ensure that all terms
proportional to $  \Sigma_{\bf x}  \sigma^0_{ij} n_{ij}  $ vanish in the
exponent of (\ref{320}). Solving them  we obtain
for $ n_j/d_j $
with $ j=1,2 $:
\begin{equation}
 \frac{n_1}{d_1} =   - \frac{\sigma^0}{2 \tilde{\mu}}        
 \quad \,, \quad
 \frac{n_2}{d_2}  =    \nu  \, \frac{\sigma^0}{2 \tilde{\mu}}   \label{350}
              \,. 
\end{equation}
These values simplify
the expression  (\ref{320})
 to
\begin{align}
 &  Z_{\rm def}  = \sum_{\cal S}
   \sum_{n_{ij} \in {\cal S}}  \exp\Bigg[-\frac{H_{\rm def}[n]}{k_B T}
 \Bigg]
               \exp\Bigg[-\frac{\tilde{H}^2_{\sigma^0}}{k_B T}
  \Bigg]    .   \label{360}
\end{align}

Finally we calculate $ Z_{\rm def} $ for $ \sigma^0 =0 $.
Taking into account only the dominant defect
configurations $ n_{ij}({\bf x}) \in\{ \pm \delta_{i,1} \delta_{j,1}
\delta_{{\bf x}, {\bf x}_0},
\pm \delta_{i,2} \delta_{j,2} \delta_{{\bf x}, {\bf x}_0},
\pm \delta_{i,1} \delta_{j,2}
\delta_{{\bf x}, {\bf x}_0} \} $, we obtain as in Refs. \cite{GFCM2, Dietel1}
\begin{equation}
  Z_{\rm def}(\sigma^0=0)
\approx  \exp[2 \exp(-6.3 \, \tilde{\beta})
 +4 \exp(-13.7 \, \tilde{\beta} )]  \,.
  \label{380}
\end{equation}

From the considerations above we conclude
that the cracking transition
is in fact not a true phase transition in the Villain model
but rather a crossover. The crossover temperature can be
obtained by the assumption that the defect configurations
$ \tilde{n}_{11} $ or $ \tilde{n}_{22} $ cover
the whole crystal area at cracking, meaning that $ d_i=1 $.
A defect configuration which covers only half of the crystal
is shown in  Fig.~3a,  where $ n_{ij}= \delta_{i,1}
\delta_{j,1} \tilde{n}_{11} $ with $ n_1=-1 $ and $ d_1=2 $.
The idea behind this  crossover temperature comes from the fact
that most cracking transition models
start with a pile up of dislocations within a glide where then
by merging we obtain a cleavage dislocation being the start point
of cracking  \cite{Hirth1}.
This is shown in Fig.~3(b).
\begin{figure}
\begin{center}
 \includegraphics[height=6cm,width=8cm]{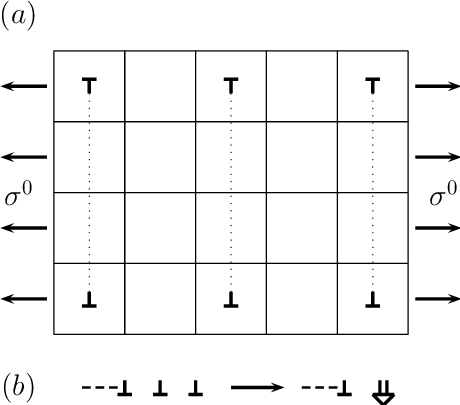}
\caption{In (a) we show a defect configuration
$ n_{ij}= \delta_{i1} \delta_{j1} \tilde{n}_{11} $ with
$ \tilde{n}_{11} $ is defined in (\ref{290}) for $ n_1=-1 $ and
$d_1=2 $. Figure (b) shows a dislocation pile up
in a glide plane leading by merging of dislocations to a
cleavage being the starting point of cracking.}
\end{center}
\end{figure}

From (\ref{350}) and the stability
criterium \cite{GFCM2} $ |\nu|\le 1 $ we obtain
that the lowest stress configuration where we have $ d_i= 1 $
is given for $ d_1=1 $ and $ n_1=1 $ resulting in a cracking stress
$ \sigma^0_b $
\begin{equation}
\frac{\sigma_b^0}{2 \tilde{\mu} } = 1 \;.        \label{385}
\end{equation}
Comparing this value with the  low-temperature
cracking stress in the mean-field approximation of the cosine-model
given by $ \sigma_b^0/ 2 \tilde{\mu} \approx 0.33 $ seen in Fig.~1
we obtain a much higher stress here.
The relative strain in the crystal phase is given by
$ \Delta u_\parallel/a =  k_B T
 \partial \ln[Z^{T \to 0}]/ \partial (N v \sigma^0) $
resulting in
\begin{equation}
 \frac{\Delta u_{\parallel}}{a} =
2 \frac{\sigma^0}{2 \tilde{\mu}}   \label{386}
\end{equation}
Comparing (\ref{386}) with the elastic part of the relative strain
$ \sigma^0/2 \tilde{\mu}  $
we obtain a factor $ 2 $ difference. By using (\ref{386}) with
(\ref{385}) we obtain a relative strain at cracking of $ 200 \% $
or $ \Delta u_\parallel/a = 2$, respectively.

Let us now address the question whether the defect
field configuration
$ \tilde{n}_{ij} $
of Eq.~(\ref{290}) is the only defect configuration
leading to the partition function
(\ref{360}). The answer is negative.
From the above derivation
we see
that
{\em any\/} $ \tilde{n}_{ij} $
arising from the substitution $ n_{ij} \rightarrow n_{ij}+
\tilde{n}_{ij} $ explained below (\ref{300}),
leads to (\ref{360}) under the condition that $ \tilde{n}_{11}({\bf k}) $
is only non-zero for $ k_2 =0 $ and $ \tilde{n}_{22}({\bf k}) $
for $ k_1 = 0 $.
All these fields will be denoted
as 
{\em defect vacuum\/}.
Such  $ \tilde{n}_{ij}({\bf x}) $
correspond to
defect stripes covering the whole
width of the crystal where the $ \tilde{n}_{ii}({\bf k}=0) $ values
are determined from the condition that the term proportional to $ \sum_{\bf x}
\sigma_{ij} n_{ij} $ vanishes in the partition function.
From the thermodynamic point of view none
of the defect configurations $ \tilde{n}_{ij}({\bf x}) $ which
fulfill the above  conditions
are preferred.
On the other hand we used the argument $ d_i=1 $ for determing the
cracking crossover stress (\ref{385})
which was justified by the defect-merging picture
in Fig.~3(b). Now suppose that we have additional
external conditions in the crystal,
for example impurities or fixed crystal defects 
generated during crystal growth, which lead to the restriction that
defects cannot cover the whole width of the system.
Then
the substitution $ n_{ij} \rightarrow n_{ij}+
\tilde{n}_{ij} $ cancels the term $ \sum_{\bf x}
\sigma_{ij} n_{ij} $ in (\ref{250}) only partly.
This is so since
the exact cancellation relies on the
fact that
the only non-vanishing contribution of (\ref{300}) in (\ref{260})
is given by the $ n_1=n_2=0 $ term, which has
 zero momentum.
Finite-length
defect stripes
have this property only approximately. These
also contribute
to (\ref{260}).
Nevertheless,
the partition function
is still approximated by (\ref{360})
if these
residual terms are
suppressed with respect to
the term 
  $ \sum_{\bf x} \sigma_{ij} n_{ij} $.
From (\ref{260})
we see
that (\ref{360}) is best fulfilled under the stripe
length restriction for that defect vacuum
$ \tilde{n}_{ij} $ which has the largest momentum region
in the vicinity of $ k_i = 0 $,
where $ n_{ii}({\bf k}\not=0 ) $ is
almost zero.
These defect configurations consists of homogeneously
distributed defect stripes in
perpendicular direction where
the density in this direction is determined from
(\ref{350}).

Note that the homogeneity can be
only be
fulfilled exactly for stresses $ \sigma_0/2 \tilde{\mu}
 \in 1 / \mathbb{N} $.
The average distance
$ d_i a $ between the stripes is then given by (\ref{350}) for $ n_i= 1 $.
The cracking condition $ d_i=1 $ is again given by the fact that the
homogeneously distributed defect stripes cover the whole crystal area
where this criterium is justified
by the defect-merging picture in Fig.~3 (b).
This leads immediately to the dimensionless cracking stress (\ref{385})
and strain (\ref{386}).

Let us finally remark
that the requirement of a homogeneously distributed
defect stripe configuration as the vacuum is also in accordance
with the conception that in real crystals the external homogeneous stress
should be relaxed homogeneously across the area by defects.

We are now prepared to
calculate the melting line by intersecting the partition functions
of the low-temperature (\ref{140}) and the high-temperature expansion
(\ref{170}). The result is
\begin{align}
& \tilde{\beta}
\exp[4 \exp(-6.3 \, \tilde{\beta} )
 + 8 \exp(-13.7 \, \tilde{\beta}) ]                \nonumber \\
&
\,~\times
\exp[-4 \exp(-5 / \tilde{\beta})]          \approx 0.81 
\label{387}
\end{align}
independent of $ \sigma^0 $.
The phase diagram
is shown in Fig.~4. It
displays
the
intersection line (\ref{387}) of
low- and high-temperature free energies ((blue) dashed curve)
and the
crossover cracking temperature (\ref{385})
plotted as
a (black) solid curve.
We obtain from Fig.~4 that the melting temperature is given by
$ 1/\tilde{\beta} \approx 1.42 $.
The melting line is given by a first order transition in the case
of a square lattice \cite{GFCM2, Dietel1}.
This was also found  in Ref.~\onlinecite{Kleinert86} by using computer
simulations.

\begin{figure}
\psfrag{x}{$1/\tilde{\beta} $}
\psfrag{y}{$\frac{\sigma^0 }{2 \tilde{\mu}}$}
\psfrag{A}{$ \frac{\sigma^0_b}{2\tilde{\mu}} $ }
\psfrag{B}{$1/\tilde{\beta}_m$}
\psfrag{R1}{${\cal R}_1$}
\psfrag{R2}{$ {\cal R}_2$}
 \psfrag{R3}{$ {\cal R}_3$}
 \includegraphics[height=5cm,width=7cm]{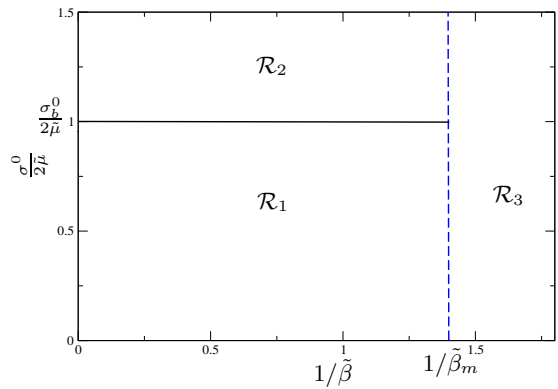}
\caption{We show the low and high-temperature intersection curve
of the free energies given by (\ref{387}) ((blue)  dashed curve).
The (black) solid curve  denotes
the cracking transition line determined by (\ref{385}).
}
\end{figure}

\section{Phase diagram of 2D-crystals }
So far we have obtained the phase diagram
in Fig.~1 by applying a mean-field approximation
in the cosine-model of crystal defect melting,
and the phase diagram in Fig.~4
from the associated Villain-type model.
The main difference lies in the fact that in the mean-field cosine-model
the cracking transition between the phases $ {\cal R}_1 $ and
$ {\cal R}_2 $ is a second order phase transition but in the
Villain-type defect model it is only a cross-over. This and the differences
in the value of the breaking stresses $ \sigma^0_b $
have their origin in calculation of the cosine-model partition function
in a mean-field approximation. It is a well-known phenomenon
in many physical systems, especially in low-dimensions,
that quantum- as well as thermodynamical fluctuations can destroy
a phase transition which appear in a mean-field approximation, leaving only
a cross-over. We have already
mentioned an example
for this
in Sect. III with the 1D XY-model \cite{Joyce1}.
Summarizing, we expect a similar phase diagram as in Fig.~4 for a real
2D crystal under stress. The stress-independent melting temperature
in this figure
is in accordance to the fact that in real physical systems there exist
a melting transition beyond cracking.
In a  triangular lattice,
 we expect
for the  melting transition line
two nearby Kosterlitz-Thouless transitions
instead of the  first-order transition
found  in the square lattice
\cite{Halperin1, Dietel1}.

In addition, our model was shown to possess
a temperature-independent cracking transition
line in Fig.~4 which is in agreement
with the experimental
determined cracking stress of 3D graphite in
Ref.~\onlinecite{Gillin1}. In that
experiment,
the cracking stress $ \sigma^0_b $ shows an anomalous temperature behaviour
only in a small temperature range  just before melting
where it starts to increase for larger temperatures.
Such an increase of the cracking stress was also seen  in other
experiments where this increase 
starts even for smaller temperatures  \cite{Malmstrom1}.
This behaviour is not fully understood yet.
That the cracking stress has a small temperature dependence is expected
and should be revealed when going beyond the elastic, lowest order 
gradient expansion
approximation used here when deriving the Villain lattice defect model.
These approximations are released by using the iV-approximation
in Sect.~III  to the Villain lattice defect model \cite{GFCM2} 
leading to Hamilton terms beyond elasticity. 
This leads us to the possibility to determine the physics of real 
crystal models which are not restricted to the elastic, lowest order 
gradient expansion with the following conclusion:
We expect a small temperature dependence 
of the cracking stress separating the phases $ {\cal R}_1 $ with 
$ {\cal R}_2 $ for realistic models. 
Nevertheless, the 
melting temperature transition separating $ {\cal R}_2 $ and 
$ {\cal R}_3 $ should not have a stress dependence also for more realistic 
models. 
Note that terms generated in the Hamiltonian
beyond elasticity
by using the iV-approximation  are in general
not directly connected to higher order terms of a certain real 
existing crystal.

In the following, we shall generalize our calculation of the last section to
crystals with stresses irrespective of direction.
In order to get the cracking stress one has to repeat the calculation
of Section IV where now  $ \tilde{n}_{ij} $ in (\ref{290})
contains additionally a term
of the form $ n_{12}\delta_{i,1}\,
\delta_{j,2}  (\delta_{x_1, a d_{12} \mathbb{Z}} +
\delta_{x_2, a d_{12} \mathbb{Z}}) $.
By carrying out the calculation  we obtain the same defect part of the
partition function $ Z_{\rm def} $ as in (\ref{360}).
The cracking transition is again determined
by the minimal stress where $ n_{i}/d_i =1 $ or  $ n_{12}/d_{12} =1 $,
respectively.
With this condition, we obtain for the cracking stress
\begin{equation}
\frac{\sigma^0_b}{2 \tilde{\mu}} =
\frac{1}{(1+\nu)}{\rm Min}
   \left\{ \begin{array}{c}
\left|\cos^2(\vartheta)- \frac{\nu}{1+\nu}
\right|^{-1} \! ,\\[0.15cm]
\left|\sin^2(\vartheta)- \frac{\nu}{1+\nu}
\right|^{-1} \!  ,\\[0.15cm]
\left|\cos(\vartheta) \sin(\vartheta) \right|^{-1}
\end{array}\!\!  \right\}  \label{410}
\end{equation}
where we took into account
 $ \sigma^0_{ij}= \sigma^0 (\cos(\vartheta),
 \sin(\vartheta)) \times (\cos(\vartheta), \sin(\vartheta)) $.
Here $ \vartheta $ is the angle between one crystal axis and the
external force.

We may now  calculate the relative strains orthogonal to the
 external force  by inserting in (\ref{24}) for
$ \sigma^0_{ij} $ an additional orthogonal auxiliary stress field in order
to calculate the orthogonal strains by differentiation.
This leads to relative strain values parallel to the external force
$ \Delta u_\parallel/a $ and orthogonal to it $ \Delta u_\perp/a $ of
\begin{eqnarray}
 \frac{ \Delta u_\parallel}{a}  & = &
2 \frac{\sigma^0}{2 \tilde{\mu}} \,,  \label{430} \\
 \frac{ \Delta u_\perp}{a}  & = & -2 \nu \, \frac{\sigma^0}{2 \tilde{\mu}}
\,.
  \label{435}
\end{eqnarray}

We show in Fig.~5 the
dimensionless cracking stress $ \sigma^0_b/2 \tilde{\mu}  $ (\ref{410})
for certain Poisson ratios $ \nu $. Together with (\ref{430})
we obtain parallel relative strains
$ \Delta u_\parallel/a \approx 200-400 \%$ where the concrete
value depends
on the Poisson ratio $ \nu $ and angle $ \vartheta $.
We note here once more that the relative strains (\ref{430}), (\ref{435})
are only valid for high temperatures or large time scales such that
activation barriers are no longer relevant in the crystal.

These large increases in the strain values at high-temperatures
are in accordance to observations
of Huang {\it et al} \cite{Huang1, Huang2} for carbon nanotubes mentioned
in the introduction of this paper. They found generally
relative elongations which are at least five times higher at cracking
than the tensile failure value at low temperature \cite{Huang3}.
\begin{figure}
\begin{center}
\psfrag{a1}{\scriptsize $ -0.5 $}
\psfrag{a2}{\scriptsize $ 0. $}
\psfrag{a3}{\scriptsize $ 0.25 $}
\psfrag{a4}{\scriptsize $ 0.5 $}
\psfrag{a5}{\scriptsize $ 0.75 $}
\psfrag{y}{$\frac{\sigma_b^0}{2 \tilde{\mu}}$}
\psfrag{x}{\hspace*{-0.5cm}$ \vartheta[{\rm degrees}]$}
 \includegraphics[height=6cm,width=8cm]{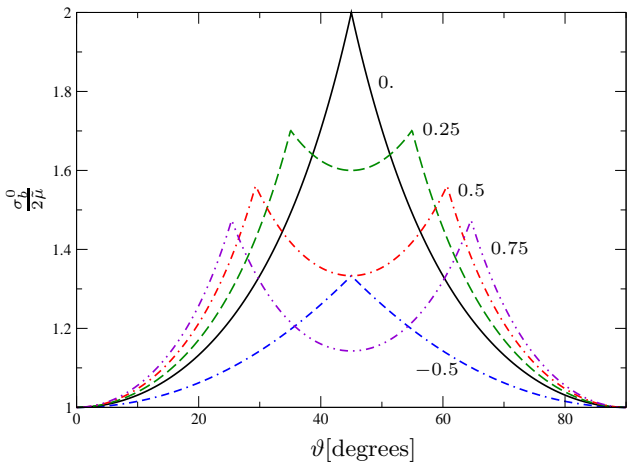}
\caption{The figure shows $ \sigma^0_b/2 \tilde{\mu} $ (\ref{410})
as a function
of the angle $ \vartheta $  between external force and the crystal
axis $ \vartheta $. The numbers at the curves denote
the Poisson ratio $ \nu $.}
\end{center}
\end{figure}

\section{Wrapped square crystals  and carbon nanotubes}

In the previous sections, we have examined the behaviour of large 
square crystals under stress. In the sequel   
we shall examine the modifications brought about   
by considering wrapped  versions of these.  
They form infinitely long thin tubes with a perimeter much larger than 
the lattice constant. Then the curve of the tube is irrelevant and 
the previous crystal model remains applicable. 

First, we shall consider {\it chiral} square tubes. 
These are defined by the property 
that the vector along the circumference of the tube lies in the direction 
of a crystal axis leading to periodic boundary conditions 
in this direction. Due to the periodic boundary conditions along the crystal
axis after integrating out the zero momentum strain fields 
described below (\ref{24}) we obtain that the mean-field 
cosine results of Sect. III as well as the results for 
the Villain model of Sects. IV and V  are also valid. 

\begin{figure}
\begin{center}
 \includegraphics[height=3.5cm,width=8cm]{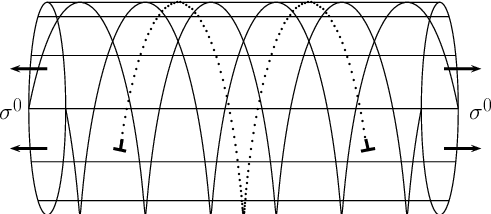}
\caption{We show a spiral-like defect configuration for a achiral 
square tube with $ n_1=-1 $ and $ d_1=2 $ by using the 
defect configuration (\ref{290}). This defect configuration is 
consistent with the 
periodic boundary conditions around the circumferent of the tube.}
\end{center}
\end{figure}

As described in the introduction section an {\it achiral tube}
is defined by the property 
that the vector along the circumference of the tube lies not 
in the direction of a crystal axis. 
Thus the periodic boundary conditions 
after the integration of the homogeneous strain fields 
are no longer in the direction of a crystal axis but in the 
circumferental direction. The mean-field 
cosine results of Sect. III are of course still
correct in this case. Also the cracking stress 
(\ref{410}) and the strain-stress  
relations (\ref{430}), (\ref{435}) are still valid. But  
the relevant defect configurations discussed below 
(\ref{386}) are no longer homogeneously distributed 
stripes but spiral-like defect stripes. One such stripe 
is shown in Fig.~6.  
By cutting the tube along the axis  
and projecting it on  the plane we can use (\ref{290})  with 
$ n_i=n_{12}=1  $
where we should 
further take into 
account $ \tilde{n}_{12} $ defined in the last section.  
This generalization becomes relevant 
for external forces not directed along a square crystal axis
valid for achiral tubes.  
The distances $ d_i $ and $ d_{12} $
between the stripes  
are determined by the periodic boundary conditions of the achiral tube.
The number of defect stripes for every sort of defects 
$ \tilde{n}_{ij} $ in axial direction is governed  
by the generalized equations of (\ref{340}) used in Section V 
if we take into account also the defect 
field $ \tilde{n}_{12} $. This means that we have to substitute in the  
generalized equation of (\ref{340}) $ n_i/d_i $ and $ n_{12}/d_{12} $ by 
the defect density along the axis which is given by 
the number of stripes of a special defect type  
divided by the number of faces along the crystal axis.  
In general this leads to the result that the homogeneous 
distributed stripes do not cover the entire  tube length.       

In the uncut tube this defect configuration 
consists of long spiral-like defect stripes whose   
extension in the axial direction is as long as possible and consistent with 
(\ref{340}). The reason lies in the fact that the substitution 
$ n_{ij} \rightarrow  n_{ij}+
\tilde{n}_{ij} $ in (\ref{250}) leads only to a cancellation of 
the external stress fields 
in the partition function when  
$ \tilde{n}_{11}({\bf k}) $ is   
zero for $ {\bf k}_2=0 $ (and similar requirements for 
$ \tilde{n}_{22}$, $ \tilde{n}_{12} $ as is outlined in Sects.~IV and V). 
Only in this case the spiral-like defect 
configurations exactly cancel the external stress field 
in the partition function (\ref{250}) as described in Sect. IV.  
Note that this requirement is exactly fulfilled only 
if a spiral-like defect covers the entire  
tube length. If a spiral-like defect stripe is 
smaller than the tube length 
where the generalized equation of (\ref{340}) applies as described in the 
last paragraph,  
the resulting stress term in the partition function (\ref{250})
leads only to negligible contributions in the free-energy density 
for infinite length and large perimeter tubes compared to the 
free energy expressions in (\ref{320}). 
 
That the  spiral-like configurations with the longest 
unbroken defect stripes which fulfill (\ref{340}) are the most 
relevant defect configurations $ \tilde{n}_{ij} $ 
is also obviously by the fact that this 
configuration has lowest energy   
$ H_{\rm def}[\tilde{n}] + H_{\sigma^0}[\tilde{n}] $ (\ref{260}), 
(\ref{270}), where the zero-momentum part (\ref{310}) is included 
in this expression.

Due to the generalized relations corresponding to (\ref{340}), 
the length of the spiral-like defect stripes increase
for increasing stresses, leading to kink propagation observed 
 by Huang {\it et al.} 
for carbon nanotubes 
\cite{Huang1, Huang2}. 
From the definition (\ref{290}) and its $ \tilde{n}_{12} $ generalization 
in Sect.~V we deduce further that the 
spiral-like defect length motion of stripes  
$ \tilde{n}_{11} $ and $\tilde{n}_{22} $ are glide motions,
whereas the stripes $ \tilde{n}_{12} $ move by climbs \cite{GFCM2}. 
Both motions were observed in the superelongation experiments 
of carbon nanotubes 
at high temperatures \cite{Huang2}. 

All this discussion leads us to the following scenario: When  
increasing the stress applied to the tube, a spiral-like defect 
for every defect type  $ \tilde{n}_i $ and $ \tilde{n}_{12} $ becomes longer 
by climb or glide, respectively.  
When one of the 
defect stripes cover the entire tube length, a new spiral-like 
defect of the same type starts to be 
formed. This goes on up to the point when the tube is covered by defects. 
This is when cracking starts.           
        
Let us end this section by discuss shortly the difference  
between real world carbon nanotubes and wrapped square crystals 
in this theoretical description. 
A carbon nanotube consist of a wrapped honeycomb lattice with two 
atoms per fundamental cell. By taking into account that the energy 
dispersion in the optical sector of the lattice displacements is negligible
in comparison to the acoustical sector \cite{Maultzsch1} leads 
to the result that it is sufficient to consider minimally 
coupled integer valued 
defect fields only in the elastic Hamiltonian of the acoustic sector 
\cite{Suzuura1}.  This then leads to a triangular lattice melting model 
since the honeycomb lattice has a triangular Bravais lattice. 
We have considered in  Ref.~\onlinecite{Dietel1} a defect
melting model for the triangular lattice (\ref{1}) for zero external stress  
in the simplest way.  
The strain-stress relations (\ref{430}), (\ref{435}) 
remain of course still valid for carbon nanotubes since  
different lattice symmetries have no influence on the long-range behaviour  
of the lattice like the 
zero momentum strain-stress relations (\ref{430}), (\ref{435}). 
Only the cracking stress 
(\ref{410}) and the defect vacuum configurations are changed.
As described above, defect configurations in a square crystal or 
its wrapped version consists  
of defect stripes which are directed along the crystal 
axes building in general spirals in achiral tubes. The same thing is  
true for the stress release in carbon nanotubes. In contrast to a square 
lattice where every vertex cuts two inequivalent lines along the crystal axes, 
every vertex cuts in a triangular lattice three inequivalent lines. This leads 
to the fact that the number of inequivalent defect stripes in the 
triangular lattice is thus a factor $ 3/2 $ larger than in the square lattice. 
By taking into account the defect-merging picture  in Fig.~3(b) for cracking 
we conclude that in carbon nanotubes the cracking stress should 
be correspondingly 
larger than in square tubes. 
The stress release now takes place on a larger amount of different 
homogeneously distributed defect stripes.   
To be more specific, we expect 
in a first rough approximation for carbon nanotubes    
cracking stresses being in the average a factor $ 3/2 $ larger than in square 
tubes. Note that this factor is similar to the 
ratio between the melting temperature difference factor of 
square crystals and triangular 
ones \cite{Dietel1} as well between honeycomb lattices and 
square crystals \cite{Dietel2}.  
In the latter case, one has to take into account properly 
the definition of the 
elongation fields in the elastic acoustic Hamiltonian as a function of the 
atomic elongations
\cite{Suzuura1} in order to obtain a temperature 
reduction factor at melting in comparison to the triangular lattice.  

A more elaborate treatment of the cracking stress of 
triangular or honeycomb lattices needs much 
more afford which is work in progress.

\section{Summary}
Motivated by recent experiments revealing
\cite{Huang1} that carbon nanotubes under
external stress show
a strong ductile behavior at high temperatures
with extremely large relative elongations before cracking, we have
 calculated in this paper
the phase diagram, the cracking stress and relative elongations
of a 2D square crystal lattice. The results
are hoped to be applicable to
carbon nanotubes, although these
form honeycomb lattices.
By starting from a Villain-type lattice defect
model we have derived in Section II,
 using the inverse Villain
approximation, an XY-like model for crystals.
When calculating within
this model the phase diagram in mean-field approximation in the
$ (T, \sigma^0) $ plane we have obtained two phase transition lines (see Fig.~1).
We have found a second-order transition at lower temperatures
which we identified as the cracking transition
line and a vertical second-order line as the  melting transition
beyond cracking.
The dimensionless cracking stress $ \sigma_b^0/2 \tilde{\mu } $ as well as
the relative strain rates parallel to the external force
$ \Delta u_\parallel/a  $ have upper bounds of
$ \sigma_b^0/2 \tilde{\mu } \le 1/3 $ and $ \Delta u_\parallel/a
\le  2/3 $ (see Fig.~2).

Next, we have calculated the phase diagram of the full Villain model within
low- and high-temperature expansion of the free energy.
Here we have found within the low-temperature expansion that the cracking
transition is in fact not a phase transition but a crossover. The
crossover line is identified by the requirement that the ground state
defect configuration under stress should cover the whole plane of the
crystal.
Within our model, the cracking stress is independent of temperature.
By using the intersection criterion of the low and
high-temperature expansions of the free energy
we obtain a melting temperature
which is independent of the external stress.
The whole phase diagram for external forces along one crystal axis is
shown in Fig.~4.

We have deduced in Sect.~V from the considerations above 
that a crystal under stress
should show, in general, a phase diagram as in Fig.~4. The phase transition
line in the cosine-model observed within the mean-field approximation should
vanish
upon taking
fluctuations
into account, converting it into
a cross-over line  as was shown in Sect.~IV within the Villain
lattice defect model. Nevertheless, we expect in accordance with the 
iV-approximation to the Villain model, 
Nevertheless, we expect by 
going beyond the elastic, lowest order gradient expansion approximation 
used in the Villain model, a small temperature dependence of the 
cracking stress, but no stress dependence of the melting transition 
temperature. This is motivated by the results for the cosine model 
of Sect.~III since the iV-approximation used to derive this model 
from the Villain model generate Hamilton terms 
beyond the elastic approximation.

Finally, we have calculated the cracking stress, and the
relative elongations
before cracking.
In Fig.~5 we have shown the
resulting dimensionless cracking stress
as a function of the angle between the external force and the crystal
axes for various Poisson ratios.
We have found dimensionless cracking stresses  $ \sigma_b^0/2 \tilde{\mu } $
 between $ 100\% $ and $ 200 \% $
 at high temperatures where potential barriers for defects are
no longer relevant.
The full relative strains $ \Delta u_\parallel/a $
in the direction of the
external force are twice as large as the elastic strain part.
The reason lies in the defect degrees of freedom
which then results in full relative strain rates
 $ \Delta u_\parallel/a $ of $ 200-400 \% $ at breaking depending
on the direction of the external force and the Poisson ratio.
The large difference in the breaking stresses and the strain rates
between the mean-field result of the cosine-model and the exact calculation
of the Villain lattice defect model is presumably due to the
mean-field approximation.

In Sect.~VI we have obtained that the cracking  stress relation (\ref{410}) 
as well as the strain-stress relations (\ref{430}), (\ref{435}) 
are also valid for 
wrapped square crystals. The defects are spiral-like for achiral 
tubes where defect glide and 
climbs are relevant in accordance to experiments. For honeycomb lattices 
or carbon nanotubes also the stress-strain relations (\ref{430}), 
(\ref{435}) are fulfilled 
but the cracking stress (\ref{410}) is now modified. We 
have argumented in a rough approximation 
that cracking stresses in carbon nanotubes 
should be in the average a factor $ 3/2 $ 
larger than in square tubes.


\begin{thebibliography}{99}

\bibitem{Nosolev1}
K.~S.~Novoselov, A.~K.~Geim, A.~K.~Geim, S.~V.~Morozov, D.~Jiang,
Y.~Zhang, S.~V.~Dubonos, I.~V.~Grogorieva, A.~A.~Firsov, Science
{\bf 306}, 666 (2004).

\bibitem{Nosolev2}
K.~S.~Novoselov,  D.~Jiang, F.~Schedin, T.~J.~Booth, V.~V.~Khotkevich,
S.~V.~Morozov, and  A.~K.~Geim, PNAS {\bf 102}, 10451 (2005).

\bibitem{Meyer1}
J.~C.~Meyer, A.~K.Geim, M.~I.~Katsnelson, K.~S.Novoselov, T.~J.~Booth,
and S.~Roth, Nature {\bf 446}, 60 (2007).

 \bibitem{Nelson1}
 D.~R.~Nelson, and L.~Peliti,
 J. Phys. (Paris) {\bf 48}, 1085 (1987).

\bibitem{Iijima1}
S.~Iijima, Nature {\bf 354}, 56 (1991).

\bibitem{Yakobson1}
B.~I.~Yakobson, M.~P.~Campbell, C.~J.~Brabec, and J.~Bernholc,
Comput. Mater. Sci. {\bf 8}, 341 (1997).

\bibitem{Haskins1}
R.~W.~Haskins, R.~S.~Maier, R.~M.~Ebeling, C.~P.~Marsh,
D.~L.~Majure, A.~J.~Bedna, C.~R.~Welch, B.~C.~Barker, and D.~T.~Wu,
J. Chem. Phys. {\bf 127}, 074708 (2007).


\bibitem{Yu1}
M.~F.~Yu, B.~S.~Files, S.~Arepalli, and R.~S.~Ruoff,
Phys. Rev. Lett. {\bf 84}, 5552 (2000).

\bibitem{Walters1}
D.~A.~Walters, L.~M.~Ericson, M.~J.~Casavant, J.~Liu, D.~T.~Colbert,
K.~A.~Smith, and R.~E.~Smalley, Appl. Phys. Lett. {\bf 74}, 3803 (1999).


\bibitem{Yu2}
M.~Yu, O.~Lourie, M.~J.~Dyer, K.~Moloni, T.~F.~Kelly, and
R.~S.~Ruoff, Science {\bf 287}, 637 (2000).


\bibitem{Huang1}
J.~Y.~Huang, S.~Chen, Z.~Q.~Wang, K.~Kempa, Y~.M~Wang, S~.H.~Jo, G.~Chen,
M.~S.~Dresselhaus, and  Z.~F.~Ren,
Nature {\bf 439}, 281 (2006).

\bibitem{Huang2}
J.~Y.~Huang, S.~Chen, Z.~F.~Ren, Z.~Q.~Wang, 
D.~Z.~Wang, M.~Vaziri, Z.~Suo, G.~Chen,
and M.~S.~Dresselhaus, Phys. Rev. Lett. {\bf 97}, 075501 (2006).

\bibitem{Ding1}
F.~Ding, K.~Jiao, M.~W.~Wu, and B.~I.~Yakobson, Phys. Rev. Lett.
{\bf 98}, 075503 (2007).

\bibitem{Tang1}
C.~Tang, W.~Guo, and C.~Chen, Phys. Rev. Lett. {\bf 100}, 175501 (2008).

\bibitem{Zhang2}
P.~Zhang, P.~E.~Lammert, and V.~H.~Crespi, Phys. Rev. Lett. {\bf 81}, 5346
(1998).



\bibitem{Yakobson2}
B.~I.~Yakobson, Appl. Phys. Lett. {\bf 72}, 918 (1998).

\bibitem{Nardelli1}
M.~BuongiornoNardelli, B.~I.~Yakobson, and J~Bernolc,
Phys. Rev. Lett. {\bf 81}, 4656, (1998).

\bibitem{Bosovic1}
D.~Bozovic, M.~Bockrath, J.~H.~Hafner, C.~M.~Leiber,
H.~Park, and M.~Tinkham, Phys. Rev. B {\bf 67}, 033407, (2003).

 \bibitem{Zhao1}
Q.~Zhao, M.~BuongiornoNardelli, and J.~Bernolz,
Phys. Rev. B {\bf 65}, 144105 (2002).

 \bibitem{Dumitrica1}
T.~Dumitric\v{a}, M.~Hua, and B.~I.~Yakobson, PNAS {\bf 103}, 6105 (2006).

\bibitem{Dumitrica2}
T.~Dumitric\v{a} and B.~Yakobson, Appl. Phys. Lett. {\bf 84}, 918 (2004).

\bibitem{GFCM2} H.~Kleinert, {\it Gauge Fields in Condensed Matter}, Vol. II
{\it Stresses and Defects: Differential Geometry, Crystal Melting}, World
Scientific, Singapore, 1989
(readable online at {\tt www.physik.fu-berlin.de/\~{}kleinert/re.html\#b2}).

\bibitem{Dietel1}
J.~Dietel and  H.~Kleinert, Phys. Rev. B {\bf 73}, 024113 (2006).

\bibitem{Nelson2}
D.~R.~Nelson,  in {\it Statistical Mechanics of Membranes and Surfaces},
Eds.  D.~Nelson, T.~Piran, and S.~Weinberg, World Scientific,
Singapore (2004).

\bibitem{GFCM1} H. Kleinert,{\it Gauge Fields in Condensed Matter}, Vol. I
{\it Superflow and Vortex lines: Disorder Fields, Phase Transition}, World
Scientific, Singapore, 1989 (readable online at {\tt www.physik.fu-berlin.de/\~{}kleinert/re.html\#b1}).

\bibitem{Joyce1}
G.~S.~Joyce, Phys. Rev. Lett. {\bf 19}, 583 (1967).

\bibitem{Hirth1}
J.~P.~Hirth and J.~Lothe, {\it Theory of Dislocations}, Toronto,
John Wiley \& Sons, 1982.

\bibitem{Kleinert86}
W.~Janke and H.~Kleinert, Phys. Lett. A {\bf 114}, 255 (1986);
W.~Janke and D.~Toussain, Phys. Lett. A {\bf 116}, 387 (1986).

\bibitem{Halperin1}
B.~I.~Halperin and D.~R.~Nelson, Phys. Rev. Lett. {\bf 41}, 121 (1978);
D.~R.~Nelson and B.~I.~Halperin, Phys. Rev. B {\bf 19}, 2457 (1979);
A.~P.~Young, Phys. Rev. B {\bf 19}, 1855 (1979).

\bibitem{Gillin1}
L.~M.~Gillin, J. Nucl. Mat. {\bf 23}, 280 (1967).

\bibitem{Malmstrom1}
C.~Malmstrom, R.~Keen and L.~Green,
J. Appl. Phys. {\bf 22}, 593 (1951);
W.~V.~Kotlensky, H.~E.~Martens, Nature {\bf 206}, 1246 (1965).

\bibitem{Huang3}
J.~Y.~Huang, S.~Chen, Z.~F.~Ren, Z.~Q.~Wang, K.~Kempa, M.~J.~Naughton,
G.~Chen, and M.~S.~Dresselhaus,
Phys. Rev. Lett. {\bf 98}, 185501 (2007).

\bibitem{Maultzsch1}
J.~Maultzsch, S.~Reich, C.~Thomsen, H.~Requardt, and P.~Ordej\'{o}n, 
Phys. Rev. Lett. {\bf 92}, 075501 (2004).  

\bibitem{Suzuura1} 
H.~Suzuura and T.~Ando, 
Phys. Rev. B {\bf 65}, 235412 (2002). 

\bibitem{Dietel2}
J.~Dietel and H.~Kleinert, Phys. Rev. B {\bf 79}, 075412 (2009).


\end{thebibliography}
\end{document}